\documentclass[aps,prd,nofootinbib,floatfix,preprintnumbers,twocolumn]{revtex4}

\usepackage{latexsym}
\usepackage{amsfonts,amsmath,amssymb}

\def\eq#1{{Eq.~(\ref{#1})}}
\newcommand{\LL}{Lanczos-Lovelock}

\begin{document}

\title{Entropy density of spacetime and thermodynamic interpretation of field equations of gravity in any diffeomorphism invariant theory}

\author{T. Padmanabhan}
\email{paddy@iucaa.ernet.in}

\affiliation{IUCAA, Post Bag 4, \\Ganeshkhind, Pune  411 007, India}

\begin{abstract}
I argue that the field equations of {\it any} theory of gravity which is diffeomorphism invariant must be expressible as
a thermodynamic identity,  $TdS=dE$ around any event in the spacetime. This fact can be demonstrated explicitly (and rather easily) if: (a) one  accepts  the Noether current of the theory as providing the definition for \textit{local} entropy density and (b) one is allowed to introduce the local notions of a  Rindler frame, acceleration horizon and a  Killing vector (related to translation in Rindler time) around any event. It is conceptually incorrect  --- in general ---- to invert this argument and obtain the field equations of the theory from the thermodynamic identity. I  discuss under what conditions this may be possible. Several subtleties related to these arguments are described.
\end{abstract}

\maketitle

Ever since the work of Bekenstein, Hawking, Davies and Unruh, \cite{allfour} it was known that there is an intriguing connection between  thermodynamics and physics of horizons. Recent work has further shown that, in a wide class of spacetimes both in general relativity and in  Lanczos-Lovelock theories, the field equations can be expressed as a thermodynamic identity near the horizon \cite{paddy2,AseemSudipta}. On the other hand, there has  been  an earlier attempt \cite{einsteineqnofstate} to  proceed in the \textit{opposite} direction  and obtain Einstein's field equations  from a thermodynamic identity.  This task was also attempted more recently \cite{PaddyAseem,paddyinside} for \LL\ models by arguing  that gravity is an emergent phenomenon like elasticity and its field equations must be derivable from extremising the entropy content of matter added to gravity.  

All these investigations were in specific contexts and it was never quite clear why these approaches  select out certain class of models or even whether they should. We shall first elaborate on this point and explain why we can reasonably expect the field equations to  be expressible as a thermodynamic identity for {\it any theory which obeys diffeomorphism invariance}, if the entropy $S$ in the equation $TdS=dE$ is appropriately interpreted.

 We begin by noting that the principle of equivalence suggests that (i) gravity should be interpreted as geometry in terms of a metric and that (ii) gravity affects the propagation of light. From these two facts it follows that strong gravitational fields can distort the light cone structure of the spacetime in  any metric theory of gravity.
Consider now a diffeomorphism invariant theory of gravity with some field equations $2E_{ab}=T_{ab}$ which possesses, say, black hole solutions with a horizon. (Most theories do, for reasons mentioned above; so this is far from a restrictive assumption). We now consider a physical process which allows some  matter with energy \textit{and} entropy (`cup of tea')  to fall into the horizon \cite{wheeler}.  We know that, the second law of thermodynamics will be violated for people who stay outside the horizon, unless there exists an entropy function $S(E)$ for the black hole which makes sure that the black hole gains both energy  {\it and entropy}  in the process, obeying the condition $TdS=dE$. Therefore, in \textit{any} such metric theory of gravity possessing horizons, not just in Einstein's theory, we should be able to find a function $S$ such that this result holds on-shell, for solutions with horizons. 

Of course, we know this is true. Wald \cite{Wald,Iyer} has already shown that one can construct an $S$ for which $TdS=dE$ will hold for \textit{on-shell solutions} of field equations in any diffeomorphism invariant theory and the above argument shows why the existence of such a function is inevitable. We also know that, $S$ can be taken to be the Noether charge associated with the diffeomorphism invariance of the theory and can be expressed as:
\begin{equation}
S_{\rm Wald} = \beta \int d^{D-1}x \left(r_{a} J^{a}\right)= \beta \int d^{D-2}\Sigma_{ab}\ J^{ab}
\label{walds}
\end{equation}
where $\beta=2\pi/\kappa, J^a\equiv\nabla_bJ^{ab}$ with $\kappa$ being the surface gravity of the horizon and the second integral is over any $(D-2)$ dimensional surface which is a spacelike cross-section of the Killing horizon.

On the other hand,  principle of general covariance implies that no observer enjoys a special status and we must interpret the theory in such a way that all observers have equal right to do physics using only the variables which (s)he can access.  This, in turn, requires us to take seriously those observers who may perceive a horizon in the spacetime (possibly) due to their state of motion. In fact, the key property of a horizon, viz. it acting as a one-way membrane, is \textit{always} an observer dependent statement. In the case of the Schwarzschild  metric, for example, we can provide a completely geometrical definition of the event horizon; but only those observers who stay at $r>2M$ will actually see it as an one-way membrane and not those who plunge into the $r<2M$ region.   So we should not make artificial, non-operational, distinctions in a generally covariant theory between certain horizons as `absolute' and some (like Rindler horizon) as 'observer dependent' but --- instead --- treat all horizons in an equal footing. This is a somewhat unconventional interpretation of general covariance but is physically well-founded. (I have described this point of view in detail in other papers of mine \cite{paddyinside,roorkee}.)

Now consider any event in the spacetime and choose a local inertial frame at that event. By boosting along one of the axes with an acceleration $\kappa$ we can introduce a local Rindler observer who perceives a local horizon. (There are subtleties in such a ``local'' definition which we will discuss at the end; right now let us assume the local physics maps to the well-known Rindler frame physics.). The fact that one has to treat all horizons on equal footing implies that a  local Rindler observer 
who sees an energy flux $dE$ crossing the horizon will expect it to be equal to  $TdS_{Wald}$ locally. \textit{If  Wald's definition of entropy works for a standard black hole solution in a particular theory, it should also satisfy the relation $TdS_{Wald}=dE$ for a local Rindler observer in  the same theory.} We know that, the real black hole solution and its horizon thermodynamics arises due to the field equations of the theory. Therefore the same field equations, when applied locally, must lead to this entropy balance condition $TdS_{Wald}=dE$ at each event for a Rindler observer. (As one will see, it is almost tautological).

Since the Wald entropy is the Noether charge obtained by integrating the Noether current suitably, we can use $J^a(x)$ to describe the local entropy balance. In particular, we can interpret $\beta_{loc}(r_aJ^a)$ to be the entropy flux at an event where $r_a$ is a normal vector to a local patch of surface and a suitable notion of a local temperature $\beta_{loc}^{-1}$ can be introduced. Based on the arguments given above, we expect this expression to match with the entropy flux of matter. If a local timelike Killing vector $\xi^a$ exists then the latter will be $\beta_{loc} T_{ab}\xi^ar^b$. So we expect field equations to imply $\beta_{loc} r_a J^a =\beta_{loc} T^{a b} r_a \xi_b$ under proper conditions. All we need is a local Killing vector $\xi^a$ and a local temperature giving $\beta_{loc}$; a local Rindler frame can supply us both. We shall now show  [see \eq{final} below], very simply,  that this relation does hold.

We will first briefly recall the derivation of Noether current \cite{Chargegb}. Consider a theory of gravity obtained from a generally covariant action principle
involving a gravitational Lagrangian $L_{grav}(R^{a}_{bcd}, g^{ab})$ which is a scalar made
from metric and curvature tensor. The total Lagrangian is the sum of $L_{grav}$
and the matter Lagrangian $L_m$. The variation of the gravitational Lagrangian
density generically leads to a surface term and hence can be expressed in the
form,
\begin{equation}
\delta(L_{grav}\sqrt{-g }) =\sqrt{-g }\left( E_{ab} \delta g^{ab} + \nabla_{a}\delta v^a\right). \label{variationL} 
\end{equation}
Under suitable boundary conditions which allow us to ignore the boundary terms (which can be nontrivial in an arbitrary theory but it is generally assumed that such theories can be defined somehow), the theory will lead to the field equation $2 E_{ab} = T_{ab}$ where $T_{ab}$ is defined through the usual relation $(1/2)T_{ab}=-(\delta A_{m}/\delta g^{ab})$. We also know that, for any Lagrangian $L_{grav}$, the functional
derivative $E_{ab}$ satisfies the generalized off-shell Bianchi identity:
\begin{equation}
\nabla_a E^{ab} = 0 \label{bianchi}
\end{equation}
We now consider the variations in $\delta g_{ab}$ which arise through the diffeomorphism $x^a \rightarrow x^a + \xi^a$. In this case, $\delta (L_{grav}\sqrt{-g} ) = -\sqrt{-g} \nabla_a (L_{grav} \xi^a)$, with $\delta g^{ab} = (\nabla^a \xi^b + \nabla^b \xi^a)$. Substituting these in \eq{variationL} and using \eq{bianchi}, we find that,
\begin{equation}
-\sqrt{-g} \nabla_a (L_{grav} \xi^a) =\sqrt{-g} \nabla_a \left(2 E^{ab} \xi_b  + \delta_{\xi} v^a \right)
\label{deltalgrav}
\end{equation}
where $\delta_{\xi}v^a$ represents the boundary term which arises for the specific variation
of the metric in the form $ \delta g^{ab} = ( \nabla^a \xi^b + \nabla^b \xi^a$). Equation~(\ref{deltalgrav}) can be expressed as a conservation law $\nabla_a J^a = 0$, for the current,
\begin{equation}
J^a \equiv \left(L_{grav}\xi^a + \delta_{\xi}v^a + 2E^{ab} \xi_b \right) 
\label{current}
\end{equation}
We stress that the conservation of this Noether current is \textit{off-shell in the sense that we have not assumed
the validity of equations of motion}. It arises purely as a consequence of diffeomorphism invariance of the gravitational action.  

For a generally covariant Lagrangian, we can write an explicit form of this Noether current. Quite generally, the boundary term can be expressed as \cite{Chargegb},
\begin{equation}
\delta v^a = \frac{1}{2} \alpha ^{a (b c)} \delta g_{ bc} + \frac{1}{2} \beta ^{a (b c)}_{d} \delta \Gamma^{d}_{bc}
\label{genv}
\end{equation}
where, we have used the notation $Q^{i j} = Q^{i j} + Q^{j i}$. The coefficient $\beta ^{a b c d}$ arises from the derivative of $L_{grav}$ with respect to $R^{ a b c d}$ and hence possess all the algebraic symmetries of curvature tensor. 
In the special case of diffeomorphisms, $x^a \rightarrow x^a + \xi^a$, the variation
$\delta_\xi v^a$ is given by \eq{genv} with:
\begin{equation}
\delta g_{a b} = -\nabla_{( a} \xi_{b)}; \quad\delta \Gamma^{d}_{bc} = 
 -\frac{1}{2} \nabla_{(b} \nabla_{c)} \xi^d 
 + \frac{1}{2} R^{d}_{(b c) m} \xi^m
\end{equation}
Using these above expressions in \eq{current}, it is possible to write an explicit expression for the current $J^a$ for any diffeomorphism invariant theory but --- interestingly enough --- we will not need its explicit form. 

Let us now consider the form of $J^{a}(x)$ at any point ${\cal P}$ around which we introduce the notion of a local Rindler horizon along the following lines:
We begin by choosing a local inertial frame (LIF) around some event $\mathcal{P}$ 
with coordinates $X^a$ such that $\mathcal{P}$ has the coordinates $X^a=0$ in the LIF.
Let $k^a$  be
 a future directed null vector at $\mathcal{P}$ and we align the coordinates of LIF
 such that it lies in the $X-T$ plane at $\mathcal{P}$. One
   can now construct, in the neighbourhood of $\mathcal{P}$
a local Rindler frame (LRF) with the coordinates $x^a$  by the usual coordinate transformations from the inertial frame to Rindler framelines,  with acceleration $\kappa$.
 Let $\xi^a$ be the (approximate) Killing vector corresponding to translation in the Rindler time such
that the vanishing of $\xi^a\xi_a \equiv -N^2$ characterizes the location of the 
local horizon $\mathcal{H}$ in the LRF. As usual, we shall do all the computation 
on a timelike surface (``stretched horizon'') infinitesimally away from $\mathcal{H}$.
On this surface  $N=\epsilon$ where $\epsilon$ is an infinitesimal constant. 
The local temperature on the stretched horizon will be $\kappa/2\pi N$ so that 
$\beta_{\rm loc} = \beta N$. 
We also assume that the approximate, local, Killing vector $\xi^a$  satisfies two conditions at ${\cal P}$: 
\begin{equation}
 \nabla_{( a} \xi_{b)} = 0;\quad  \nabla_a \nabla_b \xi_c = R_{c b a d} \xi^d
 \label{cond1}
\end{equation}
which any true Killing vector will, of course,  satisfy everywhere. 
Let $r_a$ be the spacelike unit normal to $\Sigma$, pointing in the direction of increasing $ N$. We know that as $ N\to0$ and the stretched horizon approaches the local horizon and $ N r^i$ approaches $\xi^i$.

With this background, we compute $J^a$ for the $\xi^a$ introduced above in the neighbourhood of $\mathcal{P}$. Since it is a Killing vector locally, satisfying \eq{cond1} it immediately find that $\delta g_{a b} = \delta \Gamma^{d}_{bc} =0$ giving the current as,
\begin{equation}
J^a = \left( L_{grav} \xi^a + 2 E^{a b} \xi_b \right) 
\label{current_final}
\end{equation}
 The product $r_a J^a$ for the vector $r^a$, which satisfies $\xi^ar_a=0$ on the stretched horizon,
becomes quite simple:
\begin{equation}
r_a J^a = 2 E^{a b} r_a \xi_b \qquad 
({\rm on}\; \mathcal{H}, \;{\rm if}\; r_a\xi^a=0;\ \text {off-shell})
\end{equation}
This equation is valid around the local patch in which $\xi^a$ is the approximate Killing vector.
If we were dealing with a {\it genuine solution} to the field equations which possess a horizon and a timelike Killing vector field $\xi^a$ then we would have interpreted Noether charge associated with $J^a$ as related to the entropy of the horizon. The arguments we presented in the beginning of the paper suggests that we should continue to do so for the local Rindler observer and local horizon {\it when the equations of motion hold.} Then the quantity $ \beta_{loc} r_a J^a$ has the natural interpretation as the local entropy flux density. On using the field equations $2E_{ab}=T_{ab}$, we get the entropy current to be 
\begin{equation}
\beta_{loc} r_a J^a =\beta_{loc} T^{a b} r_a \xi_b
\label{final}
\end{equation} 
In the limit of $ N\to0$, this gives a finite result, $\beta \xi_a J^a =\beta T^{a b} \xi_a \xi_b$
and shows that the matter entropy flux is precisely balanced by the gravitational entropy flux. This is equivalent to $TdS=dE$ locally. 

Thus we have shown that,  when the equations of motion holds, the change in the Wald entropy due to local energy flux obeys the same equation which we know it obeys in the case of specific solutions in the theory having horizons. Our key new ingredient in this paper is  a \textit{local} interpretation of Noether, current as entropy current; as advertised, the algebra is almost trivial once we decide to use Wald entropy.

We stress that this approach is completely equivalent to an approach in which we work with Wald entropy defined as an integral as in \eq{walds} and 
 obtain $TdS = dE $ for the integrated expressions for $S, E$ etc. In fact,
 the local definition is more in the spirit of introducing a local Rindler frame at each event. There are some well-known ambiguities in the definition of Wald entropy (which are probably more easily tackled in the integral form) but we have bypassed all that by defining a specific Noether current in \eq{current}. 

It is possible to interpret our result in very physical terms, involving a virtual displacement of the horizon which makes `cups of tea' disappear to the outside observer (see ref. \cite{paddyinside} for more details).  
An infinitesimal displacement of a local patch of the stretched horizon in the direction of $r_a$, by an infinitesimal proper distance $\epsilon$, will change the proper volume by $dV_{prop}=\epsilon\sqrt{\sigma}d^{D-2}x$ where $\sigma_{ab}$ is the metric in the transverse space.
 The flux of energy through the surface will be  $T^a_b \xi^b r_a$ and the corresponding  entropy flux
 can be obtained by multiplying the energy flux by $\beta_{\rm loc}$.  Hence
 the `loss' of matter entropy to the outside observer because the virtual displacemet of the horizon has engulfed some hot tea is 
$\delta S_m=\beta_{\rm loc}\delta E=\beta_{\rm loc} T^{aj}\xi_a r_j dV_{prop}$. Using \eq{final} we see that it matches the
corresponding  change in the gravitational entropy thereby
 showing the validity of local entropy balance for any $\beta$. 
In this limit, $\xi^i$ also goes to $\kappa \lambda k^i$ where $\lambda $ is the affine parameter associated with the null vector $k^a$ we started with  and all the reference to LRF goes away. It is clear that the properties of LRF are relevant conceptually to define the intermediate notions but the essential result which we needed was 
\eq{current_final} which has the crux of local entropy balance.

Another physical process in which the entropy balance arises is the following:  Since $J^0$ is the Noether charge density, $\delta S = \beta_{\rm loc} u_a J^a dV_{\rm prop}$ can be interpreted
as the entropy associated with a volume $dV_{\rm prop}$ as measured
by an observer with four-velocity $u^a$ where $\beta_{\rm loc} = \beta N$ is 
the local redshifted temperature and $N$ is the lapse function. (This may provide a general way of defining observer dependent entropy, which we hope to pursue in a separate paper.) If we 
consider observers moving along the orbits of the Killing vector $\xi^a$
then $u^a = \xi^a/N$ and we get
\begin{equation}
\delta S_{\rm grav} = \beta N u_a J^a dV_{\rm prop}  =\beta  [\xi_j \xi_a T^{aj} +  L  (\xi_j \xi^j)]\, dV_{\rm prop}
\end{equation} 
As one approaches the horizon, $\xi^a\xi_a\to 0$ making the second term vanish and we 
get 
\begin{equation}
\delta S_{\rm grav} = \beta  (\xi_j \xi_a T^{aj}) \, dV_{\rm prop}
\label{sgraveqn}
\end{equation}
In the same limit $\xi^j$ will become null (and proportional to the original null vector $k^j$ we started with) and for any null vector, one can interpret  the right hand side
 of \eq{sgraveqn}
 as the 
matter entropy present in a proper volume $dV_{\rm prop}$ (see e.g.,\cite{paddyinside}). So this 
equation can be again thought of as an entropy balance equation.  

There will be strong temptation at this stage to invert the argument and try to {\it derive} the field equations from the local entropy balance. Unfortunately this 
`reverse engineering' faces   some conceptual hurdles; the mathematics will go through trivially but not the logic. Let us discuss the issues involved. 

 The key point is the following: {\it If we have a justification for interpreting the expression $\beta_{loc}(r_aJ^a)$ as entropy current, independent of the field equations, then --- and only then --- can we invert the logic} and obtain the field equations from the thermodynamic identity. However, in the absence of field equations $J^a$ is just a Noether current. It can be interpreted as entropy current {\it if and only if} field equations are assumed to hold; it is in this on-shell context that Wald showed that it is  entropy. So we have no {\it independent} justification for demanding $\beta_{loc} r_aJ^a$ should be equal to matter entropy flux. Until we come up with such a justification --- without using field equations --- we can prove that "field equations imply local entropy balance at local horizons" but not "local entropy balance at local horizons imply field equations". The issue at stake is not mathematics but logic. 
 As a simple example, consider the Noether current in Einstein's theory for a Killing vector $\xi^a$, which is proportional to $R^a_b \xi^b$. It is quite difficult to interpret this conceptually as an entropy current except by studying physical processes involving black holes, say, \textit{and using field equations.}
  In the context of Einstein's theory, one can try to give other tentative arguments --- independent of field equations --- to justify why entropy is proportional to area and study the variation of area using Raychaudhury's equation as done in \cite{einsteineqnofstate}. This is, however, not possible in a general theory.

We also note that, in general, the expression for Wald entropy is not positive definite (see e.g., \cite{entneg}).
While one may be able to live with this in the on-shell context in higher dimensional gravity theories (say, by excluding the parameter range in which the entropy becomes negative), it is not obvious whether one can use an expression which is not necessarily positive definite as a \textit{fundamental} starting point in an attempt to define gravitational entropy. (There is a possibility that some of these problems may get solved in \LL\ theories, which are special but not in general.)  All these suggests that interpreting it as entropy off-shell will be a tough task \textit{but if it can be done satisfactorily, we can indeed derive field equations from a thermodynamic argument}.
 In that case, what is given above  will provide an elegant, simple and local derivation of the result. (We will obtain the  field equations with an unspecified cosmological constant arising as an integration constant --- which can be turned to a virtue; see \cite{paddygrgreview})
 Of course, one can just postulate it and `derive' the field equations but that may not provide much insight.

Note that the field equations (with an unspecified cosmological constant) can be immediately derived from an action principle
\begin{equation}
\mathcal{A}\propto \int d^Dx \sqrt{-g}[2E_{ab}-T_{ab}]n^an^b
\end{equation} 
by varying $n_a$ and demanding that $\delta \mathcal{A}=0$ for all null vectors $n_a$.  When $n_a$ is  a null  Killing vector, the gravity part of the integrand in the action reduces to $(n_aJ^a)$.
In the context of \LL\ theories with a Lagrangian $L$, it is possible to define \cite{PaddyAseem, paddyinside} another expression for gravitational entropy density in spacetime, associated with any null vector, in the form proportional to $\mathcal{L}\equiv
    P^{abcd}\nabla_cn_a\nabla_dn_b$ where $P^{abcd}\equiv (\partial L/\partial R_{abcd})$. Then, one can obtain the equations of motion of \LL\ theories by maximising the total entropy of all null vectors simultaneously. It can be easily shown that,  the entropy density of spacetime defined here  using Noether current $\xi_a J^a$ will match with $\mathcal{L}$ on $\mathcal{H}$ up to a total divergence.  Since Einstein's theory is a special case of \LL\ theories, the result also holds for Einstein's theory. This works because of the special properties of \LL\ theories, especially the fact $\nabla_a P^{abcd}=0$. It is possible that one can give independent arguments to interpret the Noether current as entropy current in the \LL\ theories rather than in the most general case. This issue is under investigation.

We shall now discuss some conceptual and technical issues    
related to the notion of a local Rindler horizon and the notion of an approximate Killing vector. It appears that these  notions are difficult to define rigourously in the Lorentzian sector of the theory (but easier to understand in the Euclidean sector). The key question is whether we can introduce a local Rindler frame with local horizon, accelerated observer etc for a sufficiently large neighbourhood around any given event. Let us briefly analyze what is involved here. 

Consider any given event in spacetime, around which a typical component of curvature tensor has the magnitude $\sim L^{-2}$. We can then introduce  Riemann normal coordinates, $X^a$, in which the metric will have a structure $g\sim \eta +(X/L)^2$. So the light cone structure (e.g the $X=T$ line in $XT$ plane) will deviate from the inertial frame structure at an accuracy of  ${\cal O}(X^2/L^2)$. We can now boost along the $X$ axis with an acceleration $\kappa$, which introduces a second length scale $\kappa^{-1}$ into the problem. One might have thought the local Rindler physics will hold as long as $L\kappa\ll 1$ which can always be achieved by choosing a sufficiently large $\kappa$. This is not completely clear in the Lorentzian sector of the theory. Making $\kappa$ arbitrarily large will make an observer move along a hyperbola which is arbitrarily close to the $X= T$. But the $X=T$ light cone structure itself  breaks down at ${\cal O}(X^2/L^2)$ {\it independent of the value chosen for $\kappa$}. An observer can be as close to the horizon as (s)he wants but the horizon itself is an approximate construct valid only to ${\cal O}(X^2/L^2)$! It is not clear how this fact affects the definition of local notions for temperature,  horizon,  Killing vector etc. and it needs further investigation. For example, it is not clear whether we need to constrain both the first   and second derivatives of $\xi^a$  by \eq{cond1} in a \textit{neighbourhood} of any event  or whether it is sufficient to do so \textit{at an event} (which is definitely possible). These conceptual issues connected with defining the local Rindler physics exists even in the case of Einstein's theory and if they can be justified in that context, it will work in the general case.
Curiously enough these problems are somewhat easier to handle in the Euclidean sector of the theory.
  Here the horizon maps to the origin of the coordinates and the stretched horizon is a circle of infinitesimal radius. The concept of local Rindler physics is better handled in the Euclidean sector because the non-compact hyperbolas ($X^2-T^2=$ constant) map to compact circles $(X^2+T_E^2=$ constant) under Euclideanisation.   
All these issues are under investigation \cite{note1}

I thank Maulik Parikh and Sudipta Sarkar for extensive discussions on a related work of theirs \cite{mpss} which prompted me also to write up these results. I also thank Dawood Kothawala
and Aseem Paranjape for useful discussions.

\end{document}